\documentclass[a4paper,12pt]{article}

\usepackage{amsmath,amssymb,array,amsfonts}
\usepackage[dvipdfmx]{graphicx}
\usepackage{comment,color}
\usepackage{geometry}
\usepackage[bf, small, hang]{caption}
\allowdisplaybreaks

\newcommand{\be}{\begin{equation}}
\newcommand{\ee}{\end{equation}}
\newcommand{\bea}{\begin{eqnarray}}
\newcommand{\eea}{\end{eqnarray}}

\makeatletter
    
    \@addtoreset{equation}{section}
\makeatother

\usepackage
[colorlinks=true,
 linkcolor=black,
 urlcolor=magenta,
 citecolor=cyan,
 hypertex]
 {hyperref}

\begin{document}
\setlength{\baselineskip}{18pt}
\begin{titlepage}

\begin{flushright}
OCU-PHYS 399 \\ 
KOBE-TH-14-03%
\end{flushright}
\vspace{1.0cm}
\begin{center}
{\Large\bf Is the 126 GeV Higgs Boson Mass Calculable \\
\vspace*{3mm}
in Gauge-Higgs Unification?} 
\end{center}
\vspace{25mm}

\begin{center}
{\large
C.S. Lim, 
Nobuhito Maru$^*$ 
and Takashi Miura$^\dagger$
}
\end{center}
\vspace{1cm}
\centerline{{\it
The Department of Mathematics, Tokyo Woman's Christian University, Tokyo 167-8585, Japan}}

\centerline{{\it
$^*$Department of Mathematics and Physics,
Osaka City University, Osaka 558-8585, Japan}} 

\centerline{{\it
$^\dagger$Department of Physics, 
Kobe University, Kobe 657-8501, Japan}} 
%
%
\vspace{2cm}
\centerline{\large\bf Abstract}
\vspace{0.5cm}

We address a question whether the recently observed Higgs mass $M_{H} = 126$ GeV, of the order of the weak scale $M_{W}$, 
 is calculable as a finite value in the scenario of gauge-Higgs unification. 
In the scenario formulated on a flat 5-dimensional space-time, the Higgs mass is calculable, 
 being protected under the quantum correction by gauge invariance, 
 though the predicted Higgs mass is generally  too small compared with $M_{W}$. In the 6-dimensional SU(3) model, however, 
 a suitable orbifolding is known to lead to a mass of the order of $M_{W}$: $M_{H} = 2M_{W}$ at the tree level, 
 which has some similarity to the corresponding prediction by the MSSM, $M_{H} \leq (\cos \beta) M_{Z}$. 

We demonstrate first by a general argument and secondly by explicit calculations that, 
 even though the quantum correction to the quartic self-coupling of the Higgs field is UV-divergent, its deviation from that of $g^{2}$ is calculable, 
 and therefore two observables, $M_{H}^{2}$ and $\Delta \equiv (\frac{M_{H}}{2M_{W}})^{2}-1$, are both calculable in the gauge-Higgs unification scenario. 
The implication of the precise value 126 GeV to the compactification scale and the bulk mass of the matter field in our model is also discussed.

\end{titlepage}




\newpage
\section{Introduction} 

The discovery of the Higgs particle was a great success of LHC experiment \cite{ATLAS, CMS}. 
We, however, should note that the long-standing problems concerning the property of Higgs and its interactions, 
 such as the hierarchy problem, are still there and we do not have any conclusive argument of the origin of the Higgs itself. 
Many of the theories of physics beyond the standard model (BSM) have been proposed in order to solve the hierarchy problem. 
At this stage, we do not know whether the discovered scalar particle is really what the standard model predicts 
or a particle some theory of BSM has in its low energy effective theory.

On the other hand, it is interesting to note that the observed Higgs mass, $M_{H}=126$ GeV, 
 seems to give us some hints on the issues discussed above. 
Namely, the Higgs mass is roughly of the order of the weak scale $M_{W}$ 
 and therefore Higgs has turned out to be relatively ``light".  
Thus, we may say that the theories predicting light Higgs are favored among proposed BSM theories, 
if they are ever realized in nature, while strongly coupled Higgs sector seems to be ruled out. 
 
The Higgs mass of ${\cal O}(M_{W})$ may also suggest that the Higgs mass is basically handled by gauge interaction. 
For instance, in MSSM the predicted Higgs mass is not far from the weak scale, 
since the Higgs quartic coupling $\lambda$ gets contribution only from gauge interaction (the D-term contribution) 
and $\lambda \sim {\cal O}(g^{2})$ at the tree level.     

We may ask a fundamental question: is it ever possible to predict the Higgs mass? 
In fact, in the SM, the Higgs mass acquires a divergent quantum correction 
and the observed Higgs mass is only realized by an adjustment of the bare Higgs mass: the origin of the hierarchy problem. 
Thus $M_{H}$ is not predictable. On the other hand, in the MSSM, for instance, $M_{H}$ is calculable as a finite value (predictable) even under the quantum correction, 
since the relation $\lambda \sim {\cal O}(g^{2})$ holds at the tree level because of the supersymmetry. 

It may be quite interesting to ask ourselves whether there exist other possibilities of BSM theories with predictable Higgs mass $M_{H}$. 
From such a point of view, in this paper we focus on another interesting scenario of BSM, 
 i.e. ``gauge-Higgs unification (GHU)". 
In the scenario of GHU, the Higgs field is identified with an extra-space component of higher dimensional gauge field. 
The scenario itself is not new \cite{1979Manton, 1979Fairlie, 1983Hosotani}, and it has been pointed out some time ago 
that the hierarchy problem is solved in this scenario thanks to the higher-dimensional gauge symmetry \cite{1998HIL}. 

Although these scenarios, MSSM and GHU, are completely independent, they have some features in common. 
First, both aim to solve the hierarchy problem relying on some symmetries. 
Secondly, also in GHU scenario the Higgs mass is basically controlled by gauge interaction, 
just because the Higgs is nothing but a gauge field to start with in this scenario. 
Thus, $M_{H}$ is calculable in the GHU. 
In fact, after \cite{1998HIL} the finiteness of the Higgs mass has been demonstrated in various types of models of GHU and even at the two loop level \cite{examples}. 

One basic problem of GHU is that the Higgs potential does not exist at the tree level in the simplest case of 5-dimensional (5D) space-time, 
as the gauge fields in general have no potential term. 
Thus $M_{H} = 0$ at the tree level. 
Even though the Higgs mass is induced at the quantum level, it is generally too small, $M_{H}^{2} = {\cal O} (\alpha M_{W}^{2})$, 
though it may be lifted once the 5D space-time is assumed to be a curved Randall-Sundrum type background \cite{RSGHU}. 
The situation may change if the number of the extra space is greater than one. 
For instance in 6D space-time, the Higgs potential gets a contribution already at the tree level 
from a term $g^{2}[A_{5}, A_{6}]^{2}$ in $F_{MN}F^{MN}$, 
where $F_{MN}$ is a field strength of the higher dimensional gauge field $A_{M} \ (M = \mu \ (\mu = 0,1,2,3), 5, 6)$ \cite{SSSW}. 
The term $g^{2}[A_{5}, A_{6}]^{2}$ provides a non-vanishing quartic self-coupling of the Higgs field, 
 unless the $A_{5}, \ A_{6}$ components of the Higgs field are proportional to each another. 
In fact, in the 6D GHU model with $T^{2}/Z_{3}$ orbifold as its extra space, 
 the quartic coupling $\lambda$ exists at the tree level, which is given in terms of the gauge coupling $g$ as  
\be
\label{1.1} 
\lambda_{tree} = \frac{1}{2}g^{2}, 
\ee 
similarly to the case of MSSM. (\ref{1.1}) in turn implies that 
\be 
\label{1.2} 
M_{H} = 2M_{W}, 
\ee
once the Higgs field acquires its VEV ($M_{H}^{2} = 2\lambda v^{2}, \ \ M_{W} = \frac{1}{2}gv \ (v:\mbox{the VEV of the Higgs field})$).
The situation is quite similar to the case of MSSM, where 
\be 
\label{1.3} 
M_{H} \leq (\cos \beta) M_{Z},  
\ee
at the tree level. 
$\beta$ is defined as the ratio of two Higgs doublet's VEVs: $\tan \beta \equiv \langle H_u \rangle/\langle H_d \rangle$.  
Thus we expect that in GHU the Higgs mass is calculable as a finite value even after the quantum correction, just as in the case of MSSM.

It is quite interesting to note that both two scenarios of BSM which aim to solve the hierarchy problem, 
MSSM and GHU, predict the Higgs mass of the order of the weak scale $M_{W}$, being consistent with the observation. 
So, a natural question to ask next is what the observed precise value of the Higgs mass, $M_{H} = 126$ GeV, implies for these scenarios. 
Note that in MSSM the observed Higgs mass is explained by choosing a suitable SUSY-breaking mass scale $M_{SUSY}$, 
though the required $M_{SUSY}$ is claimed to be a little too high from the view point of the hierarchy problem. 

Actually, the quantum correction to the Higgs mass in MSSM is much larger than we naive expect as the quantum correction: 
$M_Z + 35~{\rm GeV} \simeq 126~{\rm GeV}$, which is comparable to the weak scale itself. 
Surprisingly, in the GHU if the quantum correction of the same size is realized, 
the corrected Higgs mass happens to be just what has been observed: $2M_{W} - 35~\mbox{GeV} \simeq 126$~GeV! 
A relative sign difference of the quantum correction is expected from 
 the difference of spin statistics of the particles running inside the loop in the quantum correction, 
 i.e. stop for the case of MSSM and Kaluza-Klein (KK) top quarks for the case of GHU, for instance. 
Most probably, the relation mentioned above is just a coincidence, but this at least motivates the study of the quantum correction to the Higgs mass in the GHU. 

To be more concrete, 
what we calculate in this paper is the quantum correction to the following two observables 
which have been now completely fixed by the recent LHC experiments at CERN:  
\bea 
&&M_{H}^{2},  
\label{1.4} \\ 
&&\Delta \equiv \left( \frac{M_{H}}{2M_{W}} \right)^{2}-1.  
\label{1.5}
\eea 
Both of $M_{H}^{2}$ and $\Delta$ turn out to vanish at the tree level in our model of GHU, as is seen from (\ref{1.2}) in the case of $\Delta$ .  
This property is the consequence of the fact that in our 6D GHU model both of $M_{H}^{2}$ and $M_{W}^{2}$ are handled by a single operator, 
 i.e. the kinetic term of higher-dimensional gauge field, as is seen in (\ref{1.10}) below. 
By the reasoning given afterwards, we focus on the quantum correction to this operator, 
 neglecting higher mass dimensional (gauge invariant) operators such as $(F_{MN}F^{MN})^{2}$. 
Thus we naturally expect that even after the quantum corrections, $M_{H}^{2}$ and $\Delta$ should be UV-finite and calculable, 
 since there does not exist any local operator, which is responsible for yielding these observables, 
 though they may get finite quantum corrections due to some non-local operators, such as Wilson-loop. 
We confirm the UV-finiteness of $M_{H}^{2}$ and $\Delta$ by explicitly calculating the quantum corrections to these observables, as we will see later. 

Let us note that in MSSM, though the ratio of the quartic coupling of the Higgs to $g^{2}$ is calculable as a function of $M_{SUSY}$, 
the quadratic term of the Higgs, coming from the ``$\mu$-term" and SUSY breaking mass-squared term, 
exists already at the tree level and is not calculable, in contrast to the case of GHU.    

To see why these two quantities vanish at the tree level, 
we concentrate on the part in the lagrangian, relevant for the Higgs 
 and $W$ boson masses through the spontaneous symmetry breaking: 
\be 
\label{1.6}
-(-\mu^{2} |h_{0}|^{2} + \lambda |h_{0}|^{4}) + \kappa |h_{0}|^{2} W^{+ \mu}W^{-}_{\mu},   
\ee 
where $h_{0}$ is the electrically neutral component of the Higgs doublet. 
 By use of the coefficients $\mu^{2}, \lambda, \kappa$, two observables are expressed as 
\bea  
&&M_{H}^{2} = 2\mu^{2},  
\label{1.7} \\ 
&&\Delta = \frac{\lambda}{\kappa}-1. 
\label{1.8} 
\eea
The coefficients at the classical level, denoted as $\mu^{2}_{tree}$ etc., are known to be 
\be 
\label{1.9} 
\mu^{2}_{tree} = 0, \ \ \lambda_{tree} = \kappa_{tree} = \frac{1}{2}g^{2}.  
\ee
Hence both $M_{H}^{2}$ and $\Delta$ vanish at the tree level. 
Note that at the tree level the spontaneous symmetry breaking does not occur: $M_{H} = M_{W} = 0$, keeping the relation $M_{H} = 2M_{W}$. 
The reason why the relations in (\ref{1.9}) hold is that the coefficients $\mu^{2},  \lambda$ and $\kappa$ are all provided 
 by a single operator in the lagrangian, i.e. the kinetic term of the higher-dimensional gauge boson
\be 
\label{1.10}
-\frac{1}{2}\mbox{Tr}(F_{MN}F^{MN}) \ (M,N = \mu, 5, 6).
\ee
This operator yields the Higgs potential via $g^{2}[A_{M}, A_{N}]^{2}$ term, but only its quartic term, not a quadratic term, 
 leading to $\mu^{2}_{tree} = 0$. 
On the other hand, the relation between the coefficients of $g^{2}[A_{5}, A_{6}]^{2}$ and $g^{2}[A_{\mu}, A_{5}]^{2}$, 
$g^{2}[A_{\mu}, A_{6}]^{2}$ yields $\lambda_{tree} = \kappa_{tree}$. 

Our main purpose is to calculate the quantum corrections to $M_{H}^{2}$ and $\Delta$ and demonstrate explicitly that these two observables are in fact calculable. 
We also compare the predictions of our model with the experimental data on these two observables obtained by the recent LHC experiments 
 and will discuss whether the observed values can be accounted for by a suitable choice of the parameters of our theory, 
 such as the compactification mass scale $M_{c} \equiv 1/R$, corresponding to $M_{SUSY}$ in MSSM. 
Here $R$ is the size of $T^{2}$ of the orbifold.  

To be strict, however, 
we should note that such naive expectation may not necessarily be realized 
 in the non-renormalizable theory like higher dimensional gauge theory, 
 since operators with higher mass dimensions induced at the quantum level may also be UV-divergent. 
To be more precise, 
 e.g., $(F_{MN}F^{MN})^{2}$ having mass dimension $d = 8$ (from the viewpoint of 4D space-time) may be harmless, 
 but $(D_{L}F_{MN})(D^{L}F^{MN}) \ (d = 6)$ may be potentially dangerous having logarithmic divergence in the quantum correction to the coefficient. 
Fortunately, we readily find that the operator with $d = 6$ contributes 
 only to 6-point self-couplings of the Kaluza-Klein (KK) zero-modes of 4D Higgs and gauge bosons, 
 which do not affect the effective lagrangian given in (\ref{1.6}). 
Though these ``irrelevant" operators still may change the form of the Higgs potential, 
 we expect the contributions are relatively suppressed by higher powers of $M_{W}^{2}/M_{c}^{2}$.      

\section{The model}  
In the scenario of GHU, the gauge group should be inevitably enlarged. 
As the simplest choice we choose SU(3) as the gauge group \cite{KLY, SSS}. 
Thus, we work in the model where as the matter field scalar fields belonging to an SU(3) triplet, 
$\Phi$, are introduced in 6D space-time with $T^{2}/Z_{3}$ orbifold as the extra space. 
The torus $T^{2}$ is assumed to have the same period $2\pi R$ in both directions of two cycles. 
$Z_{3}$ is nothing but a rotation with the angle $\frac{2\pi}{3}$ 
 in the two-dimensional extra space described by the coordinates $(x_5, x_6)$. 
In this paper we aim to demonstrate that our program to predict the Higgs mass as calculable finite value works 
 by taking a toy model. 
That is the reason why we adopt scalar fields as the matter fields. 
In order to make the model realistic we are planning to introduce fermionic matter fields in future study, 
though the mechanism to get calculable $M_{W}, \ \Delta$ will not change, 
as our argumentation is based on general features of GHU, especially the higher dimensional gauge symmetry.     

The lagrangian is given as 
\be 
\label{2.1} 
{\cal L} = (D_{M} \Phi)^{\dagger}(D^{M} \Phi) - M^{2} \Phi^{\dagger}\Phi - \frac{1}{2}\mbox{Tr}(F_{MN}F^{MN}) \ \ 
\left( F_{MN} = F^{a}_{MN}T^{a}, \ \mbox{Tr}(T^{a}T^{b}) = \frac{1}{2} \delta^{ab} \right). 
\ee 
where the covariant derivative for the triplet scalar is given as 
\be 
\label{2.2} 
D_{M} = \partial_{M} + ig A_{M} \ \ (A_{M} = A^{a}_{M}T^{a}),  
\ee 
and the bulk mass $M$ is introduced in order to avoid infra-red divergence appearing 
 in the quantum correction to the coefficient $\lambda$, as we will see later.    


The $Z_{3}$-parity for the triplet scalar is assigned as follows \cite{KKM}: 
\be 
\label{2.3} 
\Phi (x, \omega z) = \Theta_{0} \Phi (x, z) \ \ (z = x_5 + ix_6, \ \omega = e^{i\frac{2\pi}{3}}), 
\ee 
where 
\be 
\label{2.4} 
\Theta_{0} = \mbox{diag}(1, 1, \omega).  
\ee 
Thus only upper two components of the triplet have KK-zero-modes, whose mode function is just a constant. 
Note that the bulk mass term in (\ref{2.1}) is $Z_{3}$ invariant. 


The 6D field $\Phi$ is expanded in terms of mode-functions as follows 
\be 
\label{2.5} 
\Phi (x, z) = \int \frac{d^{4}p}{(2\pi)^{4}} e^{ipx} \frac{1}{12^{\frac{1}{4}}\pi R} \sum_{n, m}
\begin{pmatrix}
\phi_{n,m}^{(1)}(p) f_{n,m}^{(0)}(z) \\  
\phi_{n,m}^{(2)}(p) f_{n,m}^{(0)}(z) \\  
\phi_{n,m}^{(3)}(p) f_{n,m}^{(1)}(z) \\    
\end{pmatrix}, 
\ee 
where the KK-mode-functions are given as 
\bea 
&&1: \ \ f_{n,m}^{(0)}(z) = f_{n,m}(z) + f_{n,m}(\omega z) + f_{n,m}(\bar{\omega}z), \nonumber \\ 
&&\omega: \ \  f_{n,m}^{(1)}(z) = f_{n,m}(z) + \bar{\omega}f_{n,m}(\omega z) + \omega f_{n,m}(\bar{\omega}z), \nonumber \\ 
&&\bar{\omega}: \ \ f_{n,m}^{(2)}(z) = f_{n,m}(z) + \omega f_{n,m}(\omega z) + \bar{\omega} f_{n,m}(\bar{\omega}z), \\
\label{2.6}  
\label{2.7}
&&f_{mn}(z) = \mbox{exp} \left( \frac{i}{2R} \left\{ \left( n-\frac{n+2m}{\sqrt{3}}i \right)z + c.c. \right\} \right). 
\eea 
Note that each mode function $f_{n,m}^{(0)}(z), \ f_{n,m}^{(1)}(z), \ f_{n,m}^{(2)}(z)$ has a definite eigenvalue under the $Z_{3}$ transformation, 
``$Z_{3}$-parity", $1, \ \omega, \ \bar{\omega}$, respectively. 
$f_{0,0}^{(0)}(z)$ is that for the KK-zero-mode.

\section{Background field method and mass-squared eigenvalues} 
Our purpose is to calculate the 1-loop correction to the two- and four-point functions with vanishing external momenta 
 with respect to the Higgs and $W_{\mu}^{\pm}$ fields, namely the quantum correction to the coefficients $\mu^{2}, \ \lambda$ and $\kappa$, 
 denoted by $\delta \mu^{2}, \ \delta \lambda$ and $\delta \kappa$, respectively. 
For simplicity, in this paper we focus on the quantum correction due to the scalar matter field $\Phi$.

For that purpose we use background field method, treating not only the higgs field but also $W_{\mu}^{\pm}$ as constant fields. 
We then calculate bubble diagram of the scalar field under the influence of the background fields, 
 in order to get the effective potential concerning the background fields. 
Finally, we can read off the quantum corrections $\delta \mu^{2}, \ \delta \lambda$ and $\delta \kappa$, 
 by reading off the coefficients of the relevant operators in the Taylor-expansion of the effective potential.  

The background 4D gauge field of our interest is written as 
\be 
\label{3.1} 
A^{cl}_{\mu} = 
\begin{pmatrix}
0 & \frac{W^{+}_{\mu}}{\sqrt{2}} & 0 \\ 
\frac{W^{-}_{\mu}}{\sqrt{2}} & 0 & 0 \\ 
0 & 0 & 0   
\end{pmatrix},  
\ee 
while the background 4D scalar field is written as  
\bea  
&&A^{cl}_{z} = \frac{a}{\sqrt{2}gR} 
\begin{pmatrix}
0 & 0 & 0 \\ 
0 & 0 & 0 \\ 
0 & 1 & 0   
\end{pmatrix},  \nonumber \\ 
&&A^{cl}_{\bar{z}} = (A^{cl}_{z})^{\dagger} 
= \frac{a}{\sqrt{2}gR} 
\begin{pmatrix}
0 & 0 & 0 \\ 
0 & 0 & 1 \\ 
0 & 0 & 0   
\end{pmatrix},  
\label{3.2}
\eea 
where $A_{z} \equiv \frac{1}{2}(A_{5}-iA_{6}), \ A_{\bar{z}} = A_{z}^{\dagger}$ and $a$ is a dimensionless real field defined by use of the $h^{0}$ as follows: 
\be 
\label{3.3} 
a = \frac{1}{\sqrt{2}}g|h^{0}|R. 
\ee
Note that by a suitable re-phasing $h^{0}$ can be represented by $|h^{0}|$. 

Under the presence of the background fields the bi-linear term of the scalar $\Phi$ is written as 
\bea 
(D^{cl}_{M} \Phi)^{\dagger}(D^{cl, M} \Phi) &=& - \Phi^{\dagger} D^{cl}_{M} D^{cl, M} \Phi 
= - \Phi^{\dagger} (D^{cl}_{\mu} D^{cl, \mu} - D^{cl}_{z}D^{cl, z} - D^{cl}_{\bar{z}}D^{cl, \bar{z}})\Phi \nonumber \\ 
&=& - \Phi^{\dagger} (D^{cl}_{\mu} D^{cl, \mu} - 2D^{cl}_{z}D^{cl}_{\bar{z}} - 2D^{cl}_{\bar{z}}D^{cl}_{z})\Phi,   
\label{3.4}
\eea 
where 
\be 
\label{3.5} 
D^{cl}_{\mu} = \partial_{\mu} + ig A^{cl}_{\mu}, \ \ 
D^{cl}_{z} = \partial_{z} + ig A^{cl}_{z}, \ \ 
D^{cl}_{\bar{z}} = \partial_{\bar{z}} + ig A^{cl}_{\bar{z}},  
\ee
with $\partial_{z} \equiv \frac{1}{2}(\partial_{5} - i \partial_{6})$, etc.  

Substituting (\ref{2.5}), (\ref{2.6}) in (\ref{3.4}) and by performing integral over $x^{\mu}$ and $x_{5}, \ x_{6}$ 
 together with the orthonormal conditions for the mode-functions (\ref{2.6}), 
 we realize that the successive operations of the covariant derivatives to the 4D fields with definite $(p^{\mu}, m, n)$ 
\be 
\label{3.6} 
\begin{pmatrix}
\phi_{n,m}^{(1)}(p) \\  
\phi_{n,m}^{(2)}(p) \\  
\phi_{n,m}^{(3)}(p) \\    
\end{pmatrix} 
\ee 
is equivalent to the multiplications of the following matrices: 
\bea 
&&D^{cl}_{\mu} D^{cl, \mu} =  
- \begin{pmatrix}
(p^{\mu}p_{\mu} + \frac{g^{2}}{2}W^{+\mu}W^{-}_{\mu}) & \sqrt{2}g p^{\mu}W^{+}_{\mu} & 0 \\ 
\sqrt{2}g p^{\mu}W^{-}_{\mu} & (p^{\mu}p_{\mu} + \frac{g^{2}}{2}W^{+\mu}W^{-}_{\mu}) & 0 \\ 
0 & 0 & p^{\mu}p_{\mu}     
\end{pmatrix}, \nonumber \\ 
&&D^{cl}_{z} D^{cl}_{\bar{z}} =  
- \begin{pmatrix}
\frac{1}{3R^{2}}(n^{2} + nm + m^{2}) & 0 & 0 \\ 
0 & \frac{1}{3R^{2}}(n^{2} + nm + m^{2}) & \frac{\sqrt{2}a}{4R^{2}}(n - \frac{n+2m}{\sqrt{3}}i) \\ 
0 & \frac{\sqrt{2}a}{4R^{2}}(n + \frac{n+2m}{\sqrt{3}}i) & \frac{1}{3R^{2}}(n^{2} + nm + m^{2}) + \frac{a^{2}}{2R^{2}}    
\end{pmatrix}, \nonumber \\  
&&D^{cl}_{\bar{z}} D^{cl}_{z} =  
- \begin{pmatrix}
\frac{1}{3R^{2}}(n^{2} + nm + m^{2}) & 0 & 0 \\ 
0 & \frac{1}{3R^{2}}(n^{2} + nm + m^{2}) + \frac{a^{2}}{2R^{2}} & \frac{\sqrt{2}a}{4R^{2}}(n - \frac{n+2m}{\sqrt{3}}i) \\ 
0 & \frac{\sqrt{2}a}{4R^{2}}(n + \frac{n+2m}{\sqrt{3}}i) & \frac{1}{3R^{2}}(n^{2} + nm + m^{2})     
\end{pmatrix}. 
\label{3.7}
\eea 
Namely, 
\be 
\label{3.8}  
- (D^{cl}_{\mu}D^{cl, \mu} - 2D^{cl}_{z} D^{cl}_{\bar{z}} - 2D^{cl}_{\bar{z}} D^{cl}_{z}) = p^{\mu}p_{\mu}
I_{3} 
- {\cal M}^{2},
\ee 
where $I_{3}$ is the 3$\times$3 unit matrix and 
\be 
\label{3.9}
{\cal M}^{2} \equiv 
\begin{pmatrix}
-\frac{g^{2}}{2}W^{+\mu}W^{-}_{\mu} + M^{2}_{n,m} & - \sqrt{2}g p^{\mu}W^{+}_{\mu}  & 0 \\ 
- \sqrt{2}g p^{\mu}W^{-}_{\mu} & -\frac{g^{2}}{2}W^{+\mu}W^{-}_{\mu} + M^{2}_{n,m} + \frac{a^{2}}{R^{2}} & \frac{\sqrt{2}a}{R^{2}}(n - \frac{n+2m}{\sqrt{3}}i) \\ 
0 & \frac{\sqrt{2}a}{R^{2}}(n + \frac{n+2m}{\sqrt{3}}i) & M^{2}_{n,m} + \frac{a^{2}}{R^{2}}    
\end{pmatrix}   
\ee
with 
\be 
\label{3.10} 
M^{2}_{n,m} \equiv \frac{4}{3R^{2}}(n^{2} + nm + m^{2}).
\ee 
By a suitable re-phasing of $\phi_{n,m}^{(3)}(p)$, the matrix ${\cal M}^{2}$ is brought to       
\be 
\label{3.11}
{\cal M}^{2} \equiv 
\begin{pmatrix}
-\frac{g^{2}}{2}W^{+\mu}W^{-}_{\mu} + M^{2}_{n,m} & - \sqrt{2}g p^{\mu}W^{+}_{\mu}  & 0 \\ 
- \sqrt{2}g p^{\mu}W^{-}_{\mu} & -\frac{g^{2}}{2}W^{+\mu}W^{-}_{\mu} + M^{2}_{n,m} + \frac{a^{2}}{R^{2}} & \frac{\sqrt{2}a}{R}M_{n,m} \\ 
0 & \frac{\sqrt{2}a}{R}M_{n,m} & M^{2}_{n,m} + \frac{a^{2}}{R^{2}}    
\end{pmatrix}.    
\ee 


In order to use the background field method to get the effective potential, we need three eigenvalues of the matrix ${\cal M}^{2}$. 
Since what we are interested in are the quantum corrections to the operators in (\ref{1.6}), 
 we retain only the terms up to quadratic in $W^{\pm}_{\mu}$. 
So $W^{+\mu}W^{-}_{\mu}$ and $p^{\mu}W^{+}_{\mu}$ may be treated as if they were small perturbations 
 and we can rather easily get approximated eigenvalues up to the order, by using perturbative method. 

One way to do is to write each eigenvalue as $\lambda_{i} = \lambda^{(0)}_{i} + \epsilon_{i} \ (i = 1,2,3)$, 
 where $\lambda^{(0)}_{i}$ is the eigenvalues for the vanishing $W^{\pm}_{\mu}$ 
 and the $\epsilon_{i}$ is the small perturbation of each eigenvalue, and solve for $\epsilon_{i}$ keeping only the terms up to ${\cal O}(\epsilon)$ in the equation. 
Or, we may use the well-known wisdom in the quantum mechanics to get the energy eigenvalues by use of perturbative method, 
 such as $\langle n| H'|n\rangle, \ \sum_{m \neq n} \frac{|\langle n |H'|m \rangle|^{2}}{E^{(0)}_{n}-E^{(0)}_{m}}$ 
 for the first and second orders of perturbation of energy eigenvalues. 

We have used two methods and have confirmed that the two methods give the same result. 
We will skip the detail of the derivation of the eigenvalues of ${\cal M}^{2}$ and just give the results below.
First, three eigenvalues without perturbation, $\lambda^{(0)}_{i}$, are  
\bea 
&& \lambda^{(0)}_{1} = M^{2}_{n,m}, \nonumber \\ 
&& \lambda^{(0)}_{2} = M^{2}_{n,m} + \frac{a^{2}}{R^{2}} + \frac{\sqrt{2}a}{R}M_{n,m}, \nonumber \\ 
&& \lambda^{(0)}_{3} = M^{2}_{n,m} + \frac{a^{2}}{R^{2}} - \frac{\sqrt{2}a}{R}M_{n,m}. 
\label{3.12} 
\eea 
Then the eigenvalues up to the ${\cal O}(W^{+}W^{-})$ are given as 
\bea 
\lambda_{1} &=& M^{2}_{n,m} + \frac{g^{2}|p^{\mu}W^{+}_{\mu}|^{2}}{M^{2}_{n,m} - \frac{a^{2}}{2R^{2}}} - \frac{g^{2}}{2}W^{+\mu}W^{-}_{\mu} \nonumber \\ 
&=& \lambda^{(0)}_{1} + \left( \frac{1}{\lambda^{(0)}_{1}- \lambda^{(0)}_{2}} 
+ \frac{1}{\lambda^{(0)}_{1}- \lambda^{(0)}_{3}} \right)g^{2}|p^{\mu}W^{+}_{\mu}|^{2} - \frac{g^{2}}{2}W^{+\mu}W^{-}_{\mu}, \nonumber \\ 
\lambda_{2} &=& M^{2}_{n,m} + \frac{a^{2}}{R^{2}} + \frac{\sqrt{2}a}{R}M_{n,m} 
+ \frac{g^{2}|p^{\mu}W^{+}_{\mu}|^{2}}{\frac{a^{2}}{R^{2}}+ \frac{\sqrt{2}a}{R}M_{n,m}} - \frac{g^{2}}{4}W^{+\mu}W^{-}_{\mu} \nonumber \\ 
&=& \lambda^{(0)}_{2} + \left( \frac{1}{\lambda^{(0)}_{2}- \lambda^{(0)}_{1}} \right)g^{2}|p^{\mu}W^{+}_{\mu}|^{2} - \frac{g^{2}}{4}W^{+\mu}W^{-}_{\mu}, \nonumber \\ 
\lambda_{3} &=& M^{2}_{n,m} + \frac{a^{2}}{R^{2}} - \frac{\sqrt{2}a}{R}M_{n,m}
 + \frac{g^{2}|p^{\mu}W^{+}_{\mu}|^{2}}{\frac{a^{2}}{R^{2}} - \frac{\sqrt{2}a}{R}M_{n,m}} - \frac{g^{2}}{4}W^{+\mu}W^{-}_{\mu} \nonumber \\ 
&=& \lambda^{(0)}_{3} + \left( \frac{1}{\lambda^{(0)}_{3}- \lambda^{(0)}_{1}} \right)g^{2}|p^{\mu}W^{+}_{\mu}|^{2} - \frac{g^{2}}{4}W^{+\mu}W^{-}_{\mu}. 
\label{3.13}
\eea 
\section{Quantum corrections} 
We now obtain the quantum corrections $\delta \mu^{2}, \ \delta \lambda$ and $\delta \kappa$, 
 by calculating the effective potential as a function of the background fields and reading off the suitable coefficients 
 in the Taylor expansion of the effective potential with respect to the background fields, or equivalently $a$ and $W^{\pm}$.  

The effective potential is given by the following formula: 
\be  
\label{4.1} 
V_{eff}(a, W) = \int \frac{d^{4}p_{E}}{(2\pi)^{4}} 
\sum_{n,m} [\ln (p_{E}^{2} + M^{2} + \lambda_{1}) + \ln (p_{E}^{2} + M^{2} + \lambda_{2}) + \ln (p_{E}^{2} + M^{2} + \lambda_{3})], 
\ee 
where $p_{E}$ is a Euclidean momentum and accordingly the gauge field $W^{\pm}$ should be Wick-rotated and the replacement 
\be 
\label{4.2} 
W^{+\mu}W^{-}_{\mu} \ \ \to \ \ - W^{+}\cdot W^{-}, \ \ p_{\mu}W^{+\mu} \ \ \to \ \ - p_{E}\cdot W^{+}
\ee
is understood. For instance 
\bea  
\lambda_{1} &\to& M^{2}_{n,m} + \frac{g^{2}|p_{E} \cdot W^{+}|^{2}}{M^{2}_{n,m} - \frac{a^{2}}{2R^{2}}} + \frac{g^{2}}{2}W^{+}\cdot W^{-} \nonumber \\ 
&=& \lambda^{(0)}_{1} + \left( \frac{1}{\lambda^{(0)}_{1}- \lambda^{(0)}_{2}} + \frac{1}{\lambda^{(0)}_{1}- \lambda^{(0)}_{3}} \right) 
g^{2}|p_{E} \cdot W^{+}|^{2} + \frac{g^{2}}{2}W^{+}\cdot W^{-}. 
\label{4.3} 
\eea 
To be precise, there also exists the contribution to the effective potential due to the self-interaction of 6D gauge boson $A_{M}$. 
In this paper we have ignored the contribution, since our main purpose is to demonstrate the calculability of the two observables in the simplest framework. 
Let us note that the lagrangian for the scalar matter field and that for gauge field are separately gauge invariant, 
 and the obtained result from (\ref{4.1}) does not contradict with the gauge symmetry, which plays a crucial role in our argument. 

What we are interested in are the operators 
\be 
\label{4.4}
a^{2}, \ \ a^{4}, \ \ a^{2}W^{+\mu}W^{-}_{\mu}. 
\ee 
We will discuss the quantum corrections to these operators successively below. 
\subsection{The $a^{2}$ term}  
First we calculate the $a^{2}$ term of the effective potential. 
This operator is expected to be not induced even at the quantum level at least as a local operator 
(except for the contribution due to the Wilson loop), and therefore is expected to be UV-finite. 

We set $W^{\pm} = 0$, as we are interested in the operator $a^{2}$ that does not contain the field $W$. 
Then, we find that only the terms with $\lambda_{2,3}$ in (\ref{4.1}), depending on $a$, contribute to this operator. 
Though each of $\lambda_{2,3}$ has a term linear in $M_{n,m}$, the combined contributions can be written in terms of $M_{n,m}^{2}$: 
\bea 
&&\ln (p_{E}^{2} + M^{2} + \lambda_{2}) + \ln (p_{E}^{2} + M^{2} + \lambda_{3}) \nonumber \\ 
&&= \ln \left\{ (p_{E}^{2} + M^{2} + M^{2}_{n,m})^{2}  + 2(p_{E}^{2} + M^{2})\frac{a^{2}}{R^{2}} + \frac{a^{4}}{R^{4}} \right\}.  
\label{4.5}  
\eea
Then the $a^{2}$ term in the Taylor-expansion is easily found to be 
\be 
\label{4.6} 
2\frac{p_{E}^{2} + M^{2}}{(p_{E}^{2} + M^{2} + M^{2}_{n,m})^{2}}\frac{a^{2}}{R^{2}}.  
\ee 
Thus the induced $a^{2}$ operator at the quantum level can be written as 
\be 
\label{4.7} 
2 \frac{a^{2}}{R^{2}} \int \frac{d^{4}p_{E}}{(2\pi)^{4}} \sum_{n,m} \frac{p_{E}^{2} + M^{2}}{(p_{E}^{2} + M^{2} + M^{2}_{n,m})^{2}}.  
\ee
By using formulae, 
\bea 
\frac{1}{\alpha^{2}} &=& \int_{0}^{\infty} t e^{-\alpha t} \ dt,  
\label{4.8} \\ 
\sum_{n,m} e^{-tM^{2}_{n,m}} &=& \frac{\sqrt{3}\pi R^{2}}{2} \sum_{k,l} \frac{1}{t} e^{- \frac{(\pi R)^{2}(k^{2}+kl+l^{2})}{t}} 
 \ \ (\mbox{Poisson resummation})
\label{4.9},   
\eea 
(\ref{4.7}) can be written in a form, 
\be 
\label{4.10} 
\sqrt{3} \pi a^{2} \int_{0}^{\infty} dt \int \frac{d^{4}p_{E}}{(2\pi)^{4}} (p_{E}^{2} + M^{2}) e^{-t(p_{E}^{2} + M^{2})} 
\sum_{k,l} e^{- \frac{(\pi R)^{2}(k^{2}+kl+l^{2})}{t}}. 
\ee 

In order to see whether UV-divergence is absent, 
we focus on the ``zero-winding" sector, i.e. $k = l= 0$. 
Let us note that the Poisson resummation (\ref{4.9}) is a technique to replace the summation over the KK modes $m, \ n$ 
 by the summation over the winding numbers $k, \ l$, utilizing Fourier transformation from the momentum space to the real space of 2D extra dimensions. 
The winding numbers $k, \l$ denote how many times the closed loop of Feynman diagram is wrapped around each cycle of the torus. 
Thus, the zero-winding sector $k = l = 0$ corresponds to the limit of ``decompactification" 
and has UV-divergence coming from the quantum corrections to 6D local operators, 
 while the non-zero winding sector takes into account the long distance ($\geq R$) non-local contribution and therefore is UV-finite. 

Picking up the zero-winding sector $k = l= 0$ in (\ref{4.10}), the integral over $t$ is easily done and the remaining integral is 
\be 
\label{4.11}  
\int_{0}^{\infty} dt \int \frac{d^{4}p_{E}}{(2\pi)^{4}} (p_{E}^{2} + M^{2}) e^{-t(p_{E}^{2} + M^{2})} 
= \int \frac{d^{4}p_{E}}{(2\pi)^{4}} \times 1.  
\ee 
Though (\ref{4.11}) is superficially UV-divergent, we have to be a little careful about the treatment, 
 since a momentum cutoff violates gauge symmetry. 
So we invoke dimensional regularization method, by changing $d^{4}p_{E}$ to $d^{d}p_{E}$ ($d$: space-time dimension) and taking $d \to 4$ at the final stage. 
As the matter of fact, we find that (\ref{4.11}) just vanishes, as we expected. 
Namely,  
\bea 
&&\int \frac{d^{d}p_{E}}{(2\pi)^{d}} \times 1 = \int \frac{d^{d}p_{E}}{(2\pi)^{d}} \{ \frac{p_{E}^{2}}{p_{E}^{2}+M^{2}} + \frac{M^{2}}{p_{E}^{2}+M^{2}}\} \nonumber \\ 
&&= \frac{M^{d}}{(4\pi)^{\frac{d}{2}}} \{ (\frac{d}{2})\Gamma (- \frac{d}{2}) + \Gamma (1-\frac{d}{2}) \} = 0. 
\label{4.13}   
\eea

\subsection{The $a^{4}$ term} 
We now calculate the $a^{4}$ term, in a similar way as what we took in the calculation of the quadratic term $a^{2}$. 
Again we focus on (\ref{4.5}) to get the $a^{4}$ term:   
\be 
\label{4.14} 
\frac{a^{4}}{R^{4}} \int \frac{d^{4}p_{E}}{(2\pi)^{4}} \sum_{n,m} \left\{ \frac{1}{(p_{E}^{2} + M^{2} + M^{2}_{n.m})^{2}} 
- 2 \frac{(p_{E}^{2} + M^{2})^{2}}{(p_{E}^{2} + M^{2} + M^{2}_{n.m})^{4}} \right\}. 
\ee 
By using (\ref{4.8}), (\ref{4.9}), together with 
\be 
\label{4.15} 
\frac{1}{\alpha^{4}} = \frac{1}{6} \int_{0}^{\infty} t^{3} e^{-\alpha t} \ dt,  
\ee 
(\ref{4.14}) can be put into a form 
\be 
\label{4.16} 
\frac{\sqrt{3}}{2} \pi \frac{a^{4}}{R^{2}} \int_{0}^{\infty} dt \ \int \frac{d^{4}p_{E}}{(2\pi)^{4}} e^{-t(p_{E}^{2}+M^{2})} 
\left\{ 1 -  \frac{1}{3}(p_{E}^{2} + M^{2})^{2} t^{2} \right\} \sum_{k, l} e^{- \frac{(\pi R)^{2}(k^{2}+kl+l^{2})}{t}}.  
\ee 
To see the UV-divergence, we focus on the zero-winding sector.  
Then the integral over $t$ is easily done by use of formulae 
\bea 
\int_{0}^{\infty} dt \ e^{-t(p_{E}^{2}+M^{2})} &=& \frac{1}{p_{E}^{2}+M^{2}},  
\label{4.17} \\  
\int_{0}^{\infty} dt \ e^{-t(p_{E}^{2}+M^{2})}t^{2} &=& \frac{2}{(p_{E}^{2}+M^{2})^{3}}. 
\label{4.18} 
\eea 
Namely, 
\be 
\label{4.19} 
\int_{0}^{\infty} dt \ e^{-t(p_{E}^{2}+M^{2})} \left\{ 1 -  \frac{1}{3}(p_{E}^{2} + M^{2})^{2} t^{2} \right\} 
= \frac{1}{3}\frac{1}{p_{E}^{2}+M^{2}}. 
\ee 
Thus the zero-winding sector of (\ref{4.16}) can be written as 
\be 
\label{4.20} 
\frac{\sqrt{3}}{6} \pi \frac{a^{4}}{R^{2}} \int \frac{d^{4}p_{E}}{(2\pi)^{4}} \frac{1}{p_{E}^{2}+M^{2}}.
\ee 
This time (\ref{4.20}) is apparently UV-divergent even if we utilize the dimensional regularization method: 
\be 
\label{4.21} 
\frac{\sqrt{3}}{6} \pi \frac{a^{4}}{R^{2}}\frac{\Gamma(1-\frac{d}{2})}{(4\pi)^{\frac{d}{2}}} M^{d-2} \ \ (d \to 4). 
\ee 
\subsection{The $a^{2}W^{+\mu}W^{-}_{\mu}$ term} 
The $a^{2}W^{+\mu}W^{-}_{\mu}$ term originates from $W^{+}\cdot W^{-}$ and $|p_{E} \cdot W^{+}|^{2}$ terms in the eigenvalues $\lambda_{1,2,3}$. 

We first discuss the term linear in $W^{+}\cdot W^{-}$. 
Extracting only the linear term, 
\bea 
&&\ln (p_{E}^{2} + M^{2} + \lambda_{2}) + \ln (p_{E}^{2} + M^{2} + \lambda_{3}) \nonumber \\ 
&& \ \ \to \ \ \frac{g^{2}}{4} W^{+}\cdot W^{-} \left\{ \frac{1}{p_{E}^{2}+M^{2}+\lambda^{(0)}_{2}} + \frac{1}{p_{E}^{2}+M^{2}+\lambda^{(0)}_{3}} \right\} 
\label{4.22} 
\eea 
Let us note that the second line of the above equation 
 just corresponds to the 1-loop Feynman diagram due to the 4-point vertex 
 with respect to the fields $W^{+}$, $W^{-}$ and the scalar matter fields $\phi^{(2,3)}_{n,m}(p)$ (with one propagator for the scalar fields). 

Now in (\ref{4.22}), we retain only the quadratic term in $a$: 
\bea 
&&\ln (p_{E}^{2} + M^{2} + \lambda_{2}) + \ln (p_{E}^{2} + M^{2} + \lambda_{3}) \nonumber \\ 
&& \ \ \to \ \ \frac{g^{2}}{2} \frac{a^{2}}{R^{2}} W^{+}\cdot W^{-} 
\left\{ \frac{1}{(p_{E}^{2}+M^{2}+ M^{2}_{n,m})^{2}} - 2\frac{p_{E}^{2} + M^{2}}{(p_{E}^{2}+M^{2}+M^{2}_{n,m})^{3}} \right\}.  
\label{4.23} 
\eea       
 
Secondly, we discuss the term linear in $|p_{E} \cdot W^{+}|^{2}$. 
Extracting only the linear term, 
\bea 
&&\ln (p_{E}^{2} + M^{2} + \lambda_{1}) + \ln (p_{E}^{2} + M^{2} + \lambda_{2}) + \ln (p_{E}^{2} + M^{2} + \lambda_{3}) \nonumber \\ 
&& \ \ \to g^{2}|p_{E}\cdot W^{+}|^{2} \left\{ \left( \frac{1}{\lambda^{(0)}_{1} - \lambda^{(0)}_{2}} + \frac{1}{\lambda^{(0)}_{1} - \lambda^{(0)}_{3}} \right) 
\frac{1}{p_{E}^{2}+M^{2}+\lambda^{(0)}_{1}} \right. \nonumber \\ 
&& \left.+ \frac{1}{\lambda^{(0)}_{2} - \lambda^{(0)}_{1}} \frac{1}{p_{E}^{2}+M^{2}+\lambda^{(0)}_{2}} 
+ \frac{1}{\lambda^{(0)}_{3} - \lambda^{(0)}_{1}}\frac{1}{p_{E}^{2}+M^{2}+\lambda^{(0)}_{3}} \right\} \nonumber \\ 
&& = -g^{2}|p_{E}\cdot W^{+}|^{2} \frac{1}{p_{E}^{2}+M^{2}+\lambda^{(0)}_{1}} 
\left( \frac{1}{p_{E}^{2}+M^{2}+\lambda^{(0)}_{2}} + \frac{1}{p_{E}^{2}+M^{2}+\lambda^{(0)}_{3}} \right). 
\label{4.24} 
\eea 
The last line of (\ref{4.24}) just corresponds to the 1-loop Feynman diagrams due to the 3-point vertex 
 with respect to $W^{\pm}$ and two scalar fields $\phi^{(1)}_{n,m}(p)$ and $\phi^{(2,3)}_{n,m}(p)$ (with two propagators of these scalars). 

Under $p_{E}$ integration, done later on, the following replacement can be justified: 
\be 
\label{4.25} 
|p_{E}\cdot W^{+}|^{2} \ \ \to \ \ \frac{p_{E}^{2}}{d} W^{+}\cdot W^{-}, 
\ee 
assuming dimensional regularization.  
Then, (\ref{4.24}) reduces to 
\be 
\label{4.26} 
-g^{2}W^{+}\cdot W^{-} \ \frac{p_{E}^{2}}{d} \frac{1}{p_{E}^{2}+M^{2}+\lambda^{(0)}_{1}} 
\left( \frac{1}{p_{E}^{2}+M^{2}+\lambda^{(0)}_{2}} + \frac{1}{p_{E}^{2}+M^{2}+\lambda^{(0)}_{3}} \right). 
\ee 
Again, retaining only the term quadratic in $a$, we get 
\be 
\label{4.27} 
- 2g^{2} \frac{a^{2}}{R^{2}} W^{+}\cdot W^{-} \frac{p_{E}^{2}}{d} 
\left\{ \frac{1}{(p_{E}^{2}+M^{2}+ M^{2}_{n,m})^{3}} - 2\frac{p_{E}^{2} + M^{2}}{(p_{E}^{2}+M^{2}+M^{2}_{n,m})^{4}} \right\}. 
\ee
Putting (\ref{4.23}) and (\ref{4.27}) together we get 
\be 
\label{4.28} 
g^{2} \frac{a^{2}}{R^{2}} W^{+ \mu}W^{-}_{\mu} 
\left\{ -\frac{1}{2}\frac{1}{(p_{E}^{2}+M^{2}+ M^{2}_{n,m})^{2}} + \frac{(1+\frac{2}{d})p_{E}^{2} + M^{2}}
{(p_{E}^{2}+M^{2}+ M^{2}_{n,m})^{3}} - \frac{4}{d}\frac{p_{E}^{2}(p_{E}^{2} + M^{2})}{(p_{E}^{2}+M^{2}+ M^{2}_{n,m})^{4}} \right\},  
\ee 
where a replacement $W^{+}\cdot W^{-} \ \to \ - W^{+ \mu}W^{-}_{\mu}$ has been done. 

Thus, the quantum correction to the $\frac{a^{2}}{R^{2}} W^{+ \mu}W^{-}_{\mu}$ operator can be written as 
\be 
\label{4.29} 
g^{2} \frac{a^{2}}{R^{2}} W^{+ \mu}W^{-}_{\mu} \int \frac{d^{d}p_{E}}{(2\pi)^{d}} 
\sum_{n,m} \left\{ -\frac{1}{2}\frac{1}{(p_{E}^{2}+M^{2}+ M^{2}_{n,m})^{2}} + \frac{(1+\frac{2}{d})p_{E}^{2} 
+ M^{2}}{(p_{E}^{2}+M^{2}+ M^{2}_{n,m})^{3}} - \frac{4}{d}\frac{p_{E}^{2}(p_{E}^{2} + M^{2})}{(p_{E}^{2}+M^{2}+ M^{2}_{n,m})^{4}} \right\}. 
\ee 
Using another formula 
\be 
\label{4.30} 
\frac{1}{\alpha^{3}} = \frac{1}{2} \int_{0}^{\infty} t^{2}e^{-t\alpha} dt, 
\ee 
(\ref{4.29}) can be put in a form after performing Poisson resummation,  
\bea 
&&\frac{\sqrt{3}\pi}{2}g^{2} a^{2} W^{+ \mu}W^{-}_{\mu} \int_{0}^{\infty} dt \ \int \frac{d^{d}p_{E}}{(2\pi)^{d}} e^{-t(p_{E}^{2}+M^{2})}
\left\{ -\frac{1}{2} + \frac{t}{2} \left[(1+\frac{2}{d})p_{E}^{2}+ M^{2} \right] - \frac{2t^{2}}{3d}p_{E}^{2}(p_{E}^{2} + M^{2}) \right\} \nonumber \\ 
&&\times \sum_{k,l} e^{-\frac{(\pi R)^{2}(k^{2}+kl+l^{2})}{t}} . 
\label{4.31} 
\eea 
By use of a formula 
\be 
\label{4.32} 
\int_{0}^{\infty} dt \ e^{-t(p_{E}^{2}+M^{2})} t = \frac{1}{(p_{E}^{2}+M^{2})^{2}}, 
\ee 
together with (\ref{4.17}), (\ref{4.18}), the zero-winding sector of (\ref{4.31}) turns out to take a form 
\be 
\label{4.34} 
\frac{\sqrt{3}\pi}{2}g^{2} a^{2} W^{+ \mu}W^{-}_{\mu} \int \frac{d^{d}p_{E}}{(2\pi)^{d}} \left( -\frac{1}{3d} \right) \frac{p_{E}^{2}}{(p_{E}^{2}+M^{2})^{2}}.   
\ee 
By utilizing dimensional regularization method the zero-winding sector is written as  
\be 
\label{4.35} 
- \frac{\sqrt{3}}{12}\pi g^{2} a^{2} W^{+ \mu}W^{-}_{\mu} \frac{\Gamma (1-\frac{d}{2})}{(4\pi)^{\frac{d}{2}}} M^{d-2} \ \ (d \to 4).   
\ee  

\subsection{Divergent parts of the quantum corrections}  
We have seen that at the classical level 
\be 
\label{4.37} 
\lambda_{tree} = \kappa_{tree} = \frac{1}{2}g^{2}.  
\ee 
Note that the relation $M_{H} = 2 M_{W}$ at the classical level is the consequence of the relation $\lambda_{tree} = \kappa_{tree}$. 
Now we will see whether the UV-divergent parts of $\lambda$ and $\kappa$ still preserve this relation, 
so that the deviation from the relation $M_{H} = 2 M_{W}$ can be calculated as a finite value. 

The divergent parts of $\lambda$ and $\kappa$, defined by $\delta \lambda^{div}$, $\delta \kappa^{div}$ can be easily read off 
 by replacing $a$ by $h_{0}$ in (\ref{4.21}) and (\ref{4.35}), 
 according to the relation (\ref{3.3}), and changing the overall sign 
 (the effective potential contributes to the effective lagrangian with opposite sign). 
Namely, we find 
\bea 
&& \delta \lambda^{div} = \frac{\sqrt{3}}{24}\pi g^{4} R^{2} M^{d-2} \frac{\Gamma (1 - \frac{d}{2})}{(4\pi)^{\frac{d}{2}}}, \nonumber  \\ 
&& \delta \kappa^{div} = \frac{\sqrt{3}}{24}\pi g^{4} R^{2} M^{d-2} \frac{\Gamma (1 - \frac{d}{2})}{(4\pi)^{\frac{d}{2}}}.  
\label{4.39}
\eea 
We thus find $\delta \lambda^{div} = \delta \kappa^{div}$ as we expected. 
Let us note that quantum correction $\delta \mu^{2}$ is UV-finite by itself as we have seen in (\ref{4.13}).  

\section{Calculable two observables} 
The recent LHC experiments \cite{ATLAS}, \cite{CMS} have now determined the Higgs mass as $M_{H} = 126$~GeV: 
\bea 
&& M_{H}^{2} = 126^{2}~\mbox{GeV}^{2} = 1.59 \times 10^{4}~\mbox{GeV}^{2},  
\label{5.1} \\ 
&& \left( \frac{M_{H}}{2M_{W}} \right)^{2} = \left( \frac{126}{160} \right)^{2} = 0.620 
\to \ \Delta \equiv \left(\frac{M_{H}}{2M_{W}} \right)^{2} - 1 = - 0.380.   
\label{5.2}
\eea 

A remarkable thing in our model is that both of these observables $M_{H}^{2}, \ \Delta$ are calculable 
 (as finite values without need of renormalization procedure) in terms of fundamental parameters of the theory, $R$ and $M$. 
In fact, 
\bea 
&& M_{H}^{2} = 2 \delta \mu^{2},  
\label{5.3} \\ 
&& \Delta = \frac{\lambda}{\kappa} - 1 = \frac{\frac{g^{2}}{2} + \delta \lambda}{\frac{g^{2}}{2} + \delta \kappa}
-1 \simeq \frac{2}{g^{2}}(\delta \lambda - \delta \kappa)   
\label{5.4}   
\eea 
are both finite at least at the 1-loop level, thanks to the key relation $\delta \lambda^{div} = \delta \kappa^{div}$ (see \ref{4.39}). 
We now derive the finite expressions for $M_{H}^{2}$ and $\Delta$.

The quantum corrections $\delta \mu^{2}, \ \delta \lambda, \ \delta \kappa$ are obtained from (\ref{4.10}), 
(\ref{4.16}) and (\ref{4.31}) by utilizing (\ref{3.3}). 
Namely, 
\bea 
&& \delta \mu^{2} = - \frac{\sqrt{3}}{2} \pi g^{2} R^{2} \int_{0}^{\infty} dt \int \frac{d^{4}p_{E}}{(2\pi)^{4}} (p_{E}^{2} + M^{2}) e^{-t(p_{E}^{2} + M^{2})} 
\sum_{(k,l) \neq (0,0)} e^{- \frac{(\pi R)^{2}(k^{2}+kl+l^{2})}{t}}
\label{5.5} \\ 
&& \delta \lambda = \frac{\sqrt{3}}{8} \pi g^{4} R^{2} \int_{0}^{\infty} dt \ \int \frac{d^{d}p_{E}}{(2\pi)^{d}} e^{-t(p_{E}^{2}+M^{2})} 
\left\{ 1 -  \frac{1}{3}(p_{E}^{2} + M^{2})^{2} t^{2} \right\} \sum_{k, l} e^{- \frac{(\pi R)^{2}(k^{2}+kl+l^{2})}{t}}
\label{5.6} \\ 
&& \delta \kappa = - \frac{\sqrt{3}}{4}\pi g^{4} R^{2} \int_{0}^{\infty} dt \ \int \frac{d^{d}p_{E}}{(2\pi)^{d}} e^{-t(p_{E}^{2}+M^{2})}
\left\{ -\frac{1}{2} + \frac{t}{2} \left[ \left( 1+\frac{2}{d} \right)p_{E}^{2} + M^{2} \right] - \frac{2t^{2}}{3d}p_{E}^{2}(p_{E}^{2} + M^{2}) \right\} \nonumber \\ 
&&\times \sum_{k,l} e^{-\frac{(\pi R)^{2}(k^{2}+kl+l^{2})}{t}} ,  
\label{5.7}  
\eea 
where $d = 4$ is understood for $\delta \mu^{2}$, since we know that this is UV-finite, 
 while $d$ has been left arbitrary for $\delta \lambda$ and $\delta \kappa$, since they are UV-divergent. 
The difference $\delta \lambda - \delta \kappa$ is UV-finite and is given by setting $d = 4$ as 
\bea 
&&\delta \lambda - \delta \kappa 
= \frac{\sqrt{3}}{8} \pi g^{4} R^{2} \int_{0}^{\infty} dt \ \int \frac{d^{4}p_{E}}{(2\pi)^{4}} e^{-t(p_{E}^{2}+M^{2})} \times \nonumber \\
&& \left\{ -\frac{2}{3}t^{2}(p_{E}^{2} + M^{2})^{2} + \left( \frac{1}{3}M^{2}t^{2} + \frac{3}{2}t \right)(p_{E}^{2}+M^{2}) 
- \frac{1}{2}M^{2}t \right\} \sum_{(k,l) \neq (0,0)} e^{- \frac{(\pi R)^{2}(k^{2}+kl+l^{2})}{t}}.   
\label{5.8} 
\eea

By performing the integration over $p_E$ and by changing the integration variable as $\frac{R^{2}}{t} =u$, 
the finite expressions of (\ref{5.3}) and (\ref{5.4}) are given as     
\bea 
M_{H}^{2} &=& - \frac{\sqrt{3}}{16\pi} g^{2} \frac{1}{R^{2}} \sum_{(k,l) \neq (0,0)} \int_{0}^{\infty} du (2u + \hat{M}^{2}) e^{-\frac{\hat{M}^{2}}{u}} e^{- \pi^{2}(k^{2}+kl+l^{2})u},  
\label{5.12} \\ 
\Delta &=& - \frac{\sqrt{3}}{64\pi} g^{2} \sum_{(k,l) \neq (0,0)} \int_{0}^{\infty} du \left( 1 + \frac{\hat{M}^{2}}{u} + \frac{1}{3}\frac{\hat{M}^{4}}{u^{2}} \right) 
e^{-\frac{\hat{M}^{2}}{u}} e^{- \pi^{2}(k^{2}+kl+l^{2})u}, 
\label{5.13} 
\eea 
where $\hat{M} \equiv RM$ is a dimensionless parameter. 

For a specific case of $\hat{M} = 0$, the integral over $u$ can be easily performed directly or by use of the definition of gamma functions, 
 and (\ref{5.12}) and (\ref{5.13}) reduce to simple expressions 
\bea 
M_{H}^{2} &=& - \frac{\sqrt{3}}{8\pi^{5}} g^{2} \frac{1}{R^{2}} \sum_{(k,l) \neq (0,0)} \frac{1}{(k^{2}+kl+l^{2})^{2}},  
\label{5.14} \\ 
\Delta &=& - \frac{\sqrt{3}}{64\pi^{3}} g^{2} \sum_{(k,l) \neq (0,0)} \frac{1}{k^{2}+kl+l^{2}}. 
\label{5.15} 
\eea 

Let us note (\ref{5.14}) is finite while the sum over $k, l$ in (\ref{5.15}) is divergent. 
In fact, roughly speaking in the sum $\sum_{(k,l) \neq (0,0)} \frac{1}{k^{2}+kl+l^{2}}$ the contribution from the region of large $k, l$ behaves as an integral  
\be 
\label{5.16}
\sum_{(k,l) \neq (0,0)} \frac{1}{k^{2}+kl+l^{2}} \sim \int \frac{dx_{5}dx_{6}}{x_{5}^{2} + x_{5}x_{6} + x_{6}^{2}},   
\ee 
which is logarithmically divergent (as the contribution from the region of large $x_{5,6}$). 
This logarithmic divergence comes from the region of larger $k, l$ and therefore is a sort of IR-divergence. 
In fact, we easily see that (\ref{4.14}) has an IR divergence coming from the contribution of the zero-KK-mode sector $n = m = 0$ for the case of $M = 0$, 
while (\ref{4.29}) does not have for $d = 4$. 
Thus $\delta \lambda - \delta \kappa$ should have an IR divergence.

We thus find that non-vanishing $M$ is necessary to avoid the IR divergence. 
This argument, in turn, suggests that when $\hat{M}$ is small, $\Delta$ behaves as $\propto \log \hat{M}$, 
 in order to be consistent with the logarithmic IR-divergence.

\section{The effect of the quantum correction to the kinetic term} 

In this section, for completeness, 
 we consider another possible contribution to the observable $\Delta$, i.e. the contribution of the quantum corrections of kinetic terms of the fields $W^{\pm}$ 
 and $h_{0}$ to the coefficients $\kappa$ and $\lambda$. Concerning $\mu^{2}$, the parameter is purely due to quantum effect and therefore 
 the effect of the quantum correction affects $\mu^{2}$ only at the two-loop level and can be safely ignored. 

Bare fields $W _{\mu}^{\pm}$ and $h_{0}$ are written in terms of renormalized fields $W_{r\mu}^{\pm}$ and $h_{r0}$ as 
\bea 
&&W _{\mu}^{\pm} = \sqrt{Z_W} W_{r\mu}^{\pm}, \nonumber \\ 
&&h_{0} = \sqrt{Z_{h}} h_{r0}.
\label{7.1}
\eea 
Then the formula for the quantum correction to the parameter $\Delta$ is modified from (\ref{5.4}) to the following, so that it accounts for the effect of the quantum corrections to the kinetic terms:   
\bea 
(\frac{M_{H}}{2M_{W}})^{2} &=& \frac{\lambda_{r}}{\kappa_{r}} = \frac{(\frac{g^{2}}{2} + \delta \lambda)Z_{h}^{2}}{(\frac{g^{2}}{2} + \delta \kappa)Z_{W}Z_{h}} =  
\frac{(\frac{g^{2}}{2} + \delta \lambda)(1+Z_{h}-1)}{(\frac{g^{2}}{2} + \delta \kappa)(1+Z_{W}-1)}  \nonumber \\ 
&\simeq &  1 + \frac{2}{g^{2}}(\delta \lambda - \delta \kappa) + \{(Z_{h}-1)-(Z_{W}-1) \} \nonumber \\ 
&\to & \ \ \Delta = \frac{2}{g^{2}}(\delta \lambda - \delta \kappa) + \{(Z_{h}-1)-(Z_{W}-1) \},  
\label{7.2}
\eea    
where $\lambda_{r}, \ \kappa_{r}$ denote the renormalized couplings. Thus what we should calculate is the difference of the counterterms $(Z_{h}-1) - (Z_{W}-1)$. 
We would like to point out that the difference is expected not to suffer from UV-divergence and therefore the observable $\Delta$ is still calculable. 
This is because the quantum corrections  to the kinetic terms of the Higgs and $W^{\pm}$ boson are nothing but 
 the quantum corrections to the relevant operators $F_{\mu 5}F^{\mu 5}, \ F_{\mu 6}F^{\mu 6}$ and $F_{\mu \nu}F^{\mu \nu}$. 
These two types of operators, however, are both included in a single operator $F_{MN}F^{MN}$, 
and the divergent parts of the quantum corrections to the Higgs and $W^{\pm}$ kinetic terms are expected to be the same. 
 
\subsection{Calculations of self-energy diagrams}  

\begin{figure}[htbp]
  \begin{center}
   \includegraphics[width=100mm]{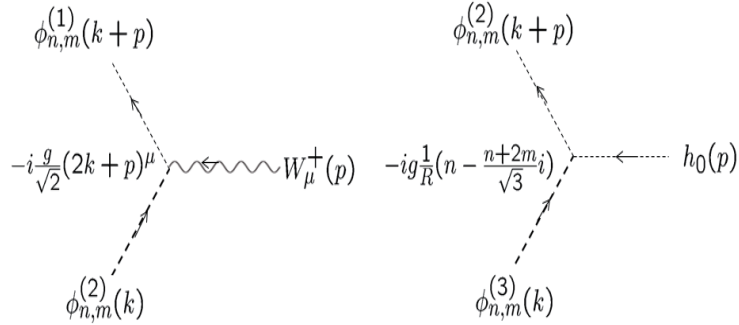}
   \end{center}
   \vspace*{-50mm}
\caption{ 
The Feynman rules for $W^{\pm}_{\mu}$ and $h_{0}$ interaction vertices. 
The pair of integers $(n, m)$ denote KK modes of scalar matter fields.}
\label{Feynmanrule}
\end{figure}

In the calculations of the quantum corrections to the kinetic terms of $W^{\pm}$ and $h_{0}$, 
 we cannot use the background field method and we just calculate each self-energy diagram according to Feynman rules. 
The derivation of the necessary Feynman rules for the 3-point vertices of $W$ and $h_{0}$ is straightforward, 
 noting $\partial_{z} f_{n,m}^{(1)}(z) = \frac{i}{2R}(n-\frac{n+2m}{\sqrt{3}}i)f_{n,m}^{(0)}(z)$, etc. 
The derived rules are given in Fig.\ref{Feynmanrule}.   

By use of the Feynman rules shown in the Fig.\ref{Feynmanrule}, self-energy diagram of $W^{\pm}_{\mu}$ with 4-momentum $p^{\mu}$ is calculated as follows, 
 by noting that the mass-squared of $\phi^{(i)}_{n,m} \ (i= 1,2,3)$ are all degenerated for vanishing background fields 
 and is given by $M^{2} + M_{n,m}^{2} \ (M_{n,m}^{2} = \frac{4}{3R^{2}}(n^{2}+nm+m^{2}))$, as is easily seen from (\ref{3.8}) and (\ref{3.11}):  
\bea
&&-i\frac{g^{2}}{2}\sum_{n,m}\int \frac{d^{d}k}{(2\pi)^{d}} \frac{(2k+p)^{\mu}(2k+p)^{\nu}}{[(k+p)^{2}-(M^{2}+M_{n,m}^{2})][k^{2}-(M^{2}+M_{n,m}^{2})]} \nonumber \\ 
&&= -i\frac{g^{2}}{2}\sum_{n,m}\int \frac{d^{d}k'}{(2\pi)^{d}} \int_{0}^{1}d\alpha 
\frac{4k'^{\mu}k'^{\nu} + (1-2\alpha)^{2}p^{\mu}p^{\nu}}{[k'^{2}+\alpha(1-\alpha)p^{2} - (M^{2}+M_{n,m}^{2})]^{2}},   
\label{7.5}
\eea
where the change of integration variable, $k \to k' = k + \alpha p$, has been done. Now we focus on the ${\cal O}(p^{2})$ terms, 
 since we are interested in the quantum correction to the kinetic term for $W^{\pm}_{\mu}$. 
 (The $p$ independent term is relevant for mass renormalization of the gauge boson 
 and should vanish when combined with the Feynman diagram due to 4-point vertex of the gauge boson and the scalar fields.) 
Then we get 
\bea 
&&-i\frac{g^{2}}{2}\sum_{n,m}\int \frac{d^{d}k'}{(2\pi)^{d}} \int_{0}^{1}d\alpha \ 
\{ \frac{-8\alpha(1-\alpha)\frac{k'^{2}}{d}g^{\mu\nu}p^{2}}{[k'^{2}- (M^{2}+M_{n,m}^{2})]^{3}} 
+ \frac{(1-2\alpha)^{2}p^{\mu}p^{\nu}}{[k'^{2}- (M^{2}+M_{n,m}^{2})]^{2}}\} \nonumber \\ 
&&= -i\frac{g^{2}}{2}\sum_{n,m}\int \frac{d^{d}k'}{(2\pi)^{d}} \{- \frac{4}{3}\frac{\frac{k'^{2}}{d}}{[k'^{2}- (M^{2}+M_{n,m}^{2})]^{3}} p^{2}g^{\mu\nu} 
+ \frac{1}{3}\frac{1}{[k'^{2}- (M^{2}+M_{n,m}^{2})]^{2}}p^{\mu}p^{\nu}\}. 
\label{7.6}
\eea  
Now let us note a useful relation easily obtained by use of dimensional regularization: 
\be 
\label{7.7}
4 \int \frac{d^{d}k'}{(2\pi)^{d}} \frac{\frac{k'^{2}}{d}}{[k'^{2}- (M^{2}+M_{n,m}^{2})]^{3}} 
= \int \frac{d^{d}k'}{(2\pi)^{d}} \frac{1}{[k'^{2}- (M^{2}+M_{n,m}^{2})]^{2}}.  
\ee
Thus, (\ref{7.6}) is neatly written as 
\be 
\label{7.8}
\frac{ig^{2}}{6} \sum_{n,m}\int \frac{d^{d}k'}{(2\pi)^{d}}\frac{1}{[k'^{2}- (M^{2}+M_{n,m}^{2})]^{2}} (p^{2}g^{\mu \nu} - p^{\mu}p^{\nu}),  
\ee
whose form is anticipated from gauge symmetry, which is present for vanishing VEV, $v = 0$.
After Wick-rotation to the Euclidean momentum $d^{d}k' = i d^{d} k_{E}, \ k'^{2} = - k_{E}^{2}$, we obtain the counterterm  
\be 
\label{7.9} 
Z_{W}-1 = - \frac{g^{2}}{6} \sum_{n,m}\int \frac{d^{d}k_{E}}{(2\pi)^{d}}\frac{1}{[k_{E}^{2}+ M^{2}+M_{n,m}^{2}]^{2}}.  
\ee
By use of (\ref{4.8}) and (\ref{4.9}), the equation (\ref{7.9}) can be rewritten as 
\be 
\label{7.10}
Z_{W}-1 = - \frac{\sqrt{3}\pi}{12}g^{2}R^{2} \int_{0}^{\infty} dt \ \int \frac{d^{d}k_{E}}{(2\pi)^{d}} e^{-t(k_{E}^{2}+M^{2})} \sum_{k,l} e^{- \frac{(\pi R)^{2}(k^{2}+kl+l^{2})}{t}}.
\ee

Similarly, the self-energy diagram of $h_{0}$ is calculated to be  
\bea
&&-ig^{2} \sum_{n,m} M_{n,m}^{2} \int \frac{d^{d}k}{(2\pi)^{d}} \frac{1}{[(k+p)^{2}-(M^{2}+M_{n,m}^{2})][k^{2}-(M^{2}+M_{n,m}^{2})]} \nonumber \\ 
&&= -ig^{2} \sum_{n,m} M_{n,m}^{2} \int \frac{d^{d}k'}{(2\pi)^{d}} \int_{0}^{1}d\alpha 
\frac{1}{[k'^{2}+\alpha(1-\alpha)p^{2} - (M^{2}+M_{n,m}^{2})]^{2}}.   
\label{7.11}
\eea
Again focusing on the ${\cal O}(p^{2})$ term, we obtain 
\bea 
&&2ig^{2}\sum_{n,m} M_{n,m}^{2} \int \frac{d^{d}k'}{(2\pi)^{d}} \int_{0}^{1}d\alpha 
\frac{\alpha(1-\alpha)}{[k'^{2}+\alpha(1-\alpha)p^{2} - (M^{2}+M_{n,m}^{2})]^{3}}p^{2} \nonumber \\ 
&&= \frac{ig^{2}}{3} \sum_{n,m} M_{n,m}^{2} \int \frac{d^{d}k'}{(2\pi)^{d}} \frac{1}{[k'^{2}+\alpha(1-\alpha)p^{2} - (M^{2}+M_{n,m}^{2})]^{3}}p^{2}. 
\label{7.12}
\eea 
Thus the counterterm for the Higgs kinetic term is given as 
\bea 
Z_{h}-1 &=& - \frac{g^{2}}{3} \sum_{n,m} M_{n,m}^{2} \int \frac{d^{d}k_{E}}{(2\pi)^{d}} \frac{1}{[k_{E}^{2} + M^{2}+M_{n,m}^{2}]^{3}} \nonumber \\ 
&=& - \frac{g^{2}}{3} \sum_{n,m} \int \frac{d^{d}k_{E}}{(2\pi)^{d}} 
\left\{ \frac{1}{[k_{E}^{2} + M^{2}+M_{n,m}^{2}]^{2}} - \frac{k_{E}^{2} + M^{2}}{[k_{E}^{2} + M^{2}+M_{n,m}^{2}]^{3}} \right\}. 
\label{7.13}
\eea
By use of (\ref{4.8}), (\ref{4.9}) and (\ref{4.30}), (\ref{7.13}) can be rewritten as 
\be 
\label{7.14}
Z_{h}-1 = - \frac{\sqrt{3}\pi}{6} g^{2}R^{2} \int_{0}^{\infty} dt \int \frac{d^{d}k_{E}}{(2\pi)^{d}} 
\left[1 - \frac{t}{2}(k_{E}^{2}+M^{2}) \right] e^{-t(k_{E}^{2}+M^{2})} \sum_{k,l} e^{- \frac{(\pi R)^{2}(k^{2}+kl+l^{2})}{t}}.
\ee
  
\subsubsection{The cancellation of UV-divergence} 

In order to extract the UV-divergent parts of $Z_{W}-1$ and $Z_{h}-1$, denoted by $(Z_{W}-1)^{div}$ and $(Z_{h}-1)^{div}$ respectively, 
 we concentrate on the ``zero-winding" sector, i.e. the sector of $k = l =0$ in (\ref{7.10}) and (\ref{7.14}). 

By taking the sector of $k = l =0$ in (\ref{7.10}),  
\bea 
(Z_{W}-1)^{div} &=& - \frac{\sqrt{3}\pi}{12}g^{2}R^{2} \int_{0}^{\infty} dt \ \int \frac{d^{d}k_{E}}{(2\pi)^{d}} e^{-t(k_{E}^{2}+M^{2})} \nonumber \\  
&=& - \frac{\sqrt{3}\pi}{12}g^{2}R^{2} \int \frac{d^{d}k_{E}}{(2\pi)^{d}} \frac{1}{k_{E}^{2}+M^{2}} 
= - \frac{\sqrt{3}}{12}\pi g^{2}R^{2} M^{d-2} \frac{\Gamma (1-\frac{d}{2})}{(4\pi)^{\frac{d}{2}}}, 
\label{7.15}
\eea 
where in the second line $t$-integral, similar to (\ref{4.17}), has been performed. 

Similarly, by taking the sector of $k = l =0$ in (\ref{7.14}), and by performing $t$-integrals, similar to (\ref{4.17}) and (\ref{4.32}),        
\bea 
(Z_{h}-1)^{div} &=& - \frac{\sqrt{3}\pi}{6} g^{2}R^{2} \int_{0}^{\infty} dt \int \frac{d^{d}k_{E}}{(2\pi)^{d}} 
\left[ 1 - \frac{t}{2}(k_{E}^{2}+M^{2}) \right] e^{-t(k_{E}^{2}+M^{2})} \nonumber \\  
&=& - \frac{\sqrt{3}\pi}{6}g^{2}R^{2} \int \frac{d^{d}k_{E}}{(2\pi)^{d}} 
\left\{ \frac{1}{k_{E}^{2}+M^{2}} - \frac{1}{2}\frac{k_{E}^{2}+M^{2}}{(k_{E}^{2}+M^{2})^{2}} \right\} \nonumber \\ 
&=& - \frac{\sqrt{3}\pi}{12}g^{2}R^{2} \int \frac{d^{d}k_{E}}{(2\pi)^{d}} \frac{1}{k_{E}^{2}+M^{2}} 
= - \frac{\sqrt{3}}{12}\pi g^{2}R^{2} M^{d-2} \frac{\Gamma (1-\frac{d}{2})}{(4\pi)^{\frac{d}{2}}}.
\label{7.16}
\eea 

Thus  
\be 
\label{7.17} 
(Z_{W}-1)^{div} = (Z_{h}-1)^{div}, 
\ee
as we anticipated from the operator analysis mentioned above. 
$\delta \lambda^{div} = \delta \kappa^{div}$, seen in (\ref{4.39}) 
 and the relation of (\ref{7.17}) we readily confirm that the UV-divergence is completely cancelled out in the prediction of the observable $\Delta$ given by (\ref{7.2}). 

Interestingly, we may also understand the UV-finiteness of the $\Delta$ from a different point of view. 
Namely, from (\ref{4.39}) and (\ref{7.16}) 
 we realize an interesting relation between the quantum corrections to the quartic Higgs self-coupling $\lambda$ and the Higgs kinetic term $Z_{h}$ 
 (``wave function renormalization"): 
\be 
\label{7.17a} 
\frac{2}{g^{2}}\delta \lambda^{div} + (Z_{h} - 1)^{div} = 0. 
\ee 
Similarly, we also realize from (\ref{4.39}) and (\ref{7.15}) that 
\be 
\label{7.17b}
\frac{2}{g^{2}}\delta \kappa^{div} + (Z_{W} - 1)^{div} = 0. 
\ee
Thus, we may understand that the finiteness of $\Delta$ defined by (\ref{7.2}) is due to the UV-finiteness shown 
 in (\ref{7.17a}) and (\ref{7.17b}) of the quantum corrections to $\lambda$ and $\kappa$ 
 when the renormalization effects of the Higgs and $W$ wave functions are taken into account.

Let us note that the UV-finiteness shown in (\ref{7.17a}) and (\ref{7.17b}) just reflects a well-known fact. 
Namely, in gauge theories even though each of the gauge field  $A_{M}$ and its gauge coupling constant $g$ gets divergent quantum correction, 
 the combined $gA_{M}$ is UV-finite and not renormalized. 
Note that in our model the Higgs is originally a gauge boson and $\lambda_{tree} = \kappa_{tree} = \frac{1}{2}g^{2}$ at the classical level 
 (see (\ref{4.37})). Thus, (\ref{7.17a}) and (\ref{7.17b}) imply that $\lambda_{tree}h_{0}^{2}$ and $\kappa_{tree}W^{+}_{\mu}W^{- \mu}$ are not renormalized. 

In fact, writing the renormalized Higgs quartic coupling $\lambda_{r}$ as  
\be 
\label{7.17'} 
\lambda_{tree} = Z_{\lambda} \lambda_{r},  
\ee 
by use of the renormalization factor $Z_{\lambda}$, the condition that $\lambda_{tree}h_{0}^{2}$ is not renormalized is written as 
\be 
\label{7.17''}
\lambda_{tree} h_{0}^{2} = Z_{\lambda}\lambda_{r} Z_{h} h_{r0}^2 = \lambda_{r}h_{r0}^2 \ \ \ \to \ \ \ Z_{\lambda} Z_{h} = 1 .
\ee 
Note that $\frac{1}{Z_{\lambda}} = (1 + \frac{2}{g^{2}}\delta \lambda) Z_{h}^{2}$. (Let us recall that $\lambda$ is the coupling of quartic Higgs interaction. 
That is why $(\sqrt{Z_{h}})^{4} = Z_{h}^{2}$ appears.) Thus the condition $Z_{\lambda} Z_{h} = 1$ of (\ref{7.17''}) means 
\be 
\label{7.17''''}
(1 + \frac{2}{g^{2}}\delta \lambda) Z_{h} \simeq 1 + \frac{2}{g^{2}}\delta \lambda + (Z_{h} - 1) = 1 \ \ \to \ \ 
\frac{2}{g^{2}}\delta \lambda + (Z_{h} - 1) = 0,  
\ee 
which is nothing but (\ref{7.17a}), as long as the UV-divergent part is concerned. A similar argument holds for the combination $\kappa_{tree}W^{+}_{\mu}W^{- \mu}$.

\subsubsection{The remaining finite contribution} 

After the cancellation of UV-divergence, the remaining finite contribution due to the quantum correction to the kinetic term reads as 
\bea 
&&(Z_{h}-1) - (Z_{W}-1) \nonumber \\ 
&&= - \frac{\sqrt{3}\pi}{12} g^{2}R^{2} \int_{0}^{\infty} dt \int \frac{d^{4}p_{E}}{(2\pi)^{4}} [1 - t(p_{E}^{2}+M^{2})]e^{-t(p_{E}^{2}+M^{2})} \sum_{(k,l) \neq (0.0)} e^{- \frac{(\pi R)^{2}(k^{2}+kl+l^{2})}{t}},  
\label{7.18} 
\eea
where $k_{E}$ has been replaced by $p_{E}$. By performing the integration over $p_{E}$ 
 and by changing the integration variable, $t \to u = \frac{R^{2}}{t}$, we get 
\be 
\label{7.19}
(Z_{h}-1) - (Z_{W}-1) = \frac{\sqrt{3}}{192 \pi} g^{2} \int_{0}^{\infty} du \ 
\left(1 + \frac{\hat{M}^{2}}{u} \right) e^{-\frac{\hat{M}^{2}}{u}} \sum_{(k,l) \neq (0.0)} e^{- \pi^{2}(k^{2}+kl+l^{2})u}.  
\ee

Adding this contribution to the previously obtained result (\ref{5.13}), we finally arrive at the complete result for $\Delta$, 
\be 
\label{7.20} 
\Delta = - \frac{\sqrt{3}}{64\pi} g^{2} \sum_{(k,l) \neq (0,0)} \int_{0}^{\infty} du 
\left(\frac{2}{3} + \frac{2}{3} \frac{\hat{M}^{2}}{u} + \frac{1}{3}\frac{\hat{M}^{4}}{u^{2}} \right) 
e^{-\frac{\hat{M}^{2}}{u}} e^{- \pi^{2}(k^{2}+kl+l^{2})u}. 
\ee   

\section{Numerical analysis of $M_{H}^{2}$ and $\Delta$} 
Though $M_{H}^{2}$ and $\Delta$ given in (\ref{5.12}) and (\ref{7.20}) are calculable as finite values, they cannot be obtained analytically. 
Thus, in this section we perform some numerical analysis. 
The purpose here is to see whether this toy model is roughly able to realize the (absolute values) of observed values 
 (\ref{5.1}) and (\ref{5.2}) for $M_{H}^{2}$ and $\Delta$ for suitable choices of $R$ and $\hat{M}$, 
 even if the signs of these two quantities cannot be correctly reproduced. 
Let us note that (\ref{5.12}) and (\ref{7.20}) are of the same sign, 
 while (\ref{5.1}) and (\ref{5.2}) tell us they have opposite signs. 

From such a point of view, 
 it may be useful to note that $|\Delta| = 0.380$ in (\ref{5.2}) is greater than what we naively expect as a 1-loop quantum correction, 
 i.e. a value roughly of the order $\alpha$, 
 while $M_{H}^{2}$ in (\ref{5.1}) can be naturally realized by a choice of the compactification scale $M_{c} = 1/R$ of the order of 1-10 TeV. 
Fortunately, we have a mechanism to realize such ``sizable" $|\Delta|$. 
Namely, because of the IR-singularity, we expect that for sufficiently small $\hat{M}$  
\be 
\label{6.1}  
|\Delta| \propto - \ln \hat{M}.   
\ee
Thus choosing suitably small $\hat{M}$ observed $\Delta$ should be realized. 
 (It is interesting to note that at least in 5D theory with orbifold compactification, 
 small ``$Z_{2}$-odd" bulk masses correspond to large fermion masses of the order of weak scale, such as the top quark mass.) 

From now on we thus assume that $\hat{M}$ is small enough and will confirm the expectation mentioned above. 
We first discuss $|\Delta|$. 

\begin{figure}[htbp]
  \begin{center}
   \includegraphics[width=70mm]{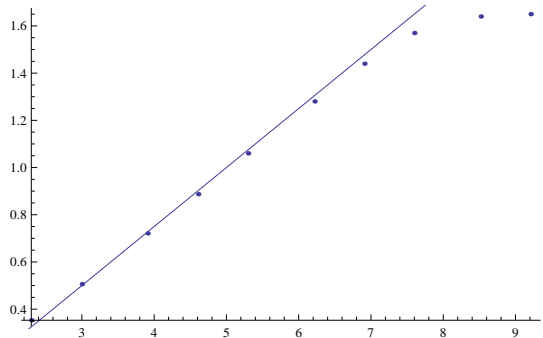}
   \end{center}
  \caption{ 
The function $\bar{F}(\hat{M}^{2})$. The horizontal axis is $- \ln \hat{M}^{2}$ and the vertical axis is $\bar{F}(\hat{M}^{2})$.    
The straight line stands for $\bar{F}(\hat{M}^{2}) = 0.25 (- \ln \hat{M}^{2})-0.25$}
  \label{Bulkmassv2}
\end{figure}

We have performed a numerical computation of the following factor in (\ref{7.20}): 
\be 
\label{6.2} 
F(\hat{M}^{2}) \equiv \sum_{(k,l) \neq (0,0)} \int_{0}^{\infty} du 
\left( \frac{2}{3} + \frac{2}{3} \frac{\hat{M}^{2}}{u} + \frac{1}{3}\frac{\hat{M}^{4}}{u^{2}} \right) e^{-\frac{\hat{M}^{2}}{u}} e^{- \pi^{2}(k^{2}+kl+l^{2})u}. 
\ee
Actually, because of the lack of the computational capability we have in our hand, we have approximated $F(\hat{M})$ by the following function 
\be 
\label{6.3} 
\bar{F}(\hat{M}^{2}) \equiv \sum_{(k,l) \neq (0,0), |k|, |l| \leq 30} \int_{0.00015}^{100} du 
\left( \frac{2}{3} + \frac{2}{3} \frac{\hat{M}^{2}}{u} + \frac{1}{3}\frac{\hat{M}^{4}}{u^{2}} \right) e^{-\frac{\hat{M}^{2}}{u}} e^{- \pi^{2}(k^{2}+kl+l^{2})u}.
\ee 
The result of the numerical calculation for the function $\bar{F}(\hat{M})$ is shown in Fig.\ref{Bulkmassv2}.  

As we expected, as $\hat{M}^{2}$ becomes small enough or equivalently as $- \ln \hat{M}^{2}$ becomes large enough, 
the function shows logarithmic behavior 
\be 
\label{6.4} 
\bar{F}(\hat{M}^{2}) \simeq 0.25 (- \ln \hat{M}^{2}). 
\ee
The reason why the function $\bar{F}(\hat{M}^{2})$ finally starts to be saturated for larger $- \ln \hat{M}^{2}$ is easily understood. 
If the sum over $k, \ l$ are taken up to arbitrarily large integers as in the original function $F(\hat{M}^{2})$, 
 the function should become arbitrarily large for sufficiently large $- \ln \hat{M}^{2}$.  
Actually, however, in the approximated function $\bar{F}(\hat{M}^{2})$, the sum over $k, \ l$ are only up to $|k| = |l| = 30$, not infinity. 
Thus the function $\bar{F}(\hat{M}^{2})$ never exceeds 
\be 
\label{6.5} 
\frac{2}{3\pi^{2}} \sum_{(k,l) \neq (0,0), |k|, |l| \leq 30} \frac{1}{k^{2}+kl+l^{2}} = 1.91, 
\ee 
which is nothing but $\bar{F}(0)$, with the region of integral being replaced by $0 \leq u < \infty$. We thus reasonably expect that the original function $F(\hat{M}^{2})$ should behave as 
\be 
\label{6.6}
F(\hat{M}^{2}) \simeq 0.25 (- \ln \hat{M}^{2}).
\ee 

We now turn to another observable $M_{H}^{2}$. In this case we can use the formula (\ref{5.14}), 
 corresponding to $\hat{M}^{2} = 0$, since the sum over $k, \ l$ is finite, in contrast to the case of $\Delta$. 
A numerical computation yields 
\be 
\label{6.7} 
\sum_{(k,l) \neq (0,0)} \frac{1}{(k^{2}+kl+l^{2})^{2}} = 7.71.
\ee

To summarize, we have obtained 
\bea 
&&|M_{H}^{2}| \simeq \frac{\sqrt{3}}{8\pi^{5}} g^{2} \frac{1}{R^{2}} \sum_{(k,l) \neq (0,0)} \frac{1}{(k^{2}+kl+l^{2})^{2}} = 0.0685 \frac{\alpha}{\sin^{2}\theta_{W}}\frac{1}{R^{2}} = 2.2 \times 10^{-3}\frac{1}{R^{2}},
\label{6.8} \\ 
&&|\Delta| = \frac{\sqrt{3}}{64\pi} g^{2} F(\hat{M}^{2}) \simeq 2.71 \times 10^{-2}\frac{\alpha}{\sin^{2}\theta_{W}}(- \ln \hat{M}^{2}) = 1.73 \times 10^{-3}(- \ln \hat{M}),   
\label{6.9} 
\eea 
where $\alpha = \frac{1}{137}$ and $\sin^{2}\theta_{W} = 0.23$ have been used. Actually, our model with SU(3) gauge group 
predicts $\sin^{2}\theta_{W} = \frac{3}{4}$, far from 0.23. 
We, however, take the observed value 0.23, hoping that in a realistic model it is realized by the introduction of brane-localized kinetic term 
or by making the gauge group semi-simple such as SU(3)$\times$U(1), SO(5)$\times$U(1).   

By comparing these results with the observed values (\ref{5.1}) and (\ref{5.2}), we finally get 
\bea 
\frac{1}{R} \simeq  2.7~\mbox{TeV}, \qquad
- \ln \hat{M} \simeq 220.   
\eea 
One problem here is that the obtained logarithmic factor (its absolute value) is ridiculously large. 
We, however, point out that when we consider the effect of heavy particle such as top quark in a realistic model, we expect to get an enhancement factor. 
Namely, concerning the top quark contribution to realize its large Yukawa coupling top quark is assigned 
 as a member of higher dimensional repr. of gauge group, e.g. 4th-rank symmetric tensor in the case of SU(3) GHU, 
 which leads to an enhancement factor in its contribution to $\Delta$ due to the group factor, basically due to the large Yukawa coupling, 
 such as $(\frac{m_{t}}{M_{W}})^{4} \sim 23$. 
Thus we expect that the problem of too large logarithmic factor or too small bulk mass may be reasonably evaded in the realistic model.          
   
\section{Summary} 
In this paper we addressed a question 
 whether the recently observed Higgs mass $M_{H} = 126$ GeV is calculable as a finite value in the scenario of gauge-Higgs unification (GHU). 
We first pointed out that the recently observed Higgs mass is of ${\cal O}(M_{W})$ 
 and seems to suggest that the Higgs mass is handled by gauge interaction, roughly speaking. 
To be more specific, we discussed that in both scenarios of GHU (formulated on 6D space-time) 
 and SUSY (MSSM) proposed mainly for the purpose of solving the hierarchy problem, 
 the quartic self-coupling of the Higgs field is governed by the gauge-principle, being ${\cal O}(g^{2})$. 
This fact led to the expectation that the deviation of the Higgs mass from the prediction at the tree level is calculable as a finite value, 
 being free from UV-divergence, in GHU scenario, not only in 5D space-time but also in higher space-time dimensions, such as 6D. 
The situation is similar to the case of MSSM, 
 where the deviation of the Higgs mass from $(\cos \beta) M_{Z}$ is calculable in terms of the SUSY breaking masses, such as the stop mass and the ``$A$-term".  

We have argued that, as a new feature of the GHU scenario, not shared by MSSM, 
not only the quartic self-coupling, but also the quadratic coupling of the Higgs is calculable as well, 
just because none of gauge invariant local operators constructed by use of higher-dimensional field strength induced at the quantum level should not have such quadratic operator for the Higgs field. 

Thus we claimed that in the GHU, as the matter of fact, we have two independent calculable observables, i.e. 
\be 
M_{H}^{2}, \ \ \ \Delta \equiv \left( \frac{M_{H}}{2M_{W}} \right)^{2}-1.
\ee 
This expectation has been confirmed by explicit calculations of the quantum corrections to these quantities in a toy model. 
Note that in our model of GHU, both quantities just vanish at the tree level. 

The model we adopted was a 6D toy model formulated on $T^{2}/Z_{3}$ orbifold as the extra-space. 
For brevity, as the matter field we introduced 6D scalar fields, behaving as an SU(3) triplet, and the quantum corrections to $M_{H}^{2}$ and $\Delta$ 
due to the self-interactions of the higher-dimensional gauge fields have not been included in our analysis. 
Note that this treatment is consistent with gauge invariance, just because each of the bi-linear terms of scalar fields and gauge 
fields has gauge symmetry independently and the quantum correction due to the each sector is gauge invariant by itself. 

Although the toy model is sufficient for the purpose to demonstrate that we have two calculable observables in the GHU scenario, 
 obviously to get the realistic values for these quantities it is necessary to work in a realistic model 
 with quarks and leptons and to incorporate the contributions due to the self-interactions of higher-dimensional gauge bosons $A_{M}$. 
We hope that the problem of the mutual sign in the quantum corrections to $M_{H}^{2}$ and $\Delta$ pointed out in this paper is solved 
 by the calculations in such realistic framework. 
We would like to report on the results of the calculations in a future publication. 

A comment on brane localized ``tadpole" terms is now in order. As discussed in \cite{SSSW}, 
 brane localized term like $F_{56}$ is allowed in a gauge invariant way if a $U(1)$ is included in the gauge group unbroken at the orbifold fixed points. 
This term, if exists, yields the Higgs mass-squared term through the commutator $g[A_{5}, A_{6}]$ in the field strength localized at the fixed points 
 whose coefficient is divergent in general, thus spoiling the calculability of the Higgs mass. 
 
We, however, would like to point out that in our model we do not suffer from this problem. 
Let us note that if the localized tadpole term ever exists with a divergent coefficient being proportional to a $\delta$-function localized at one of the fixed points, 
 it will cause a UV-divergence in the effective 4D theory, which is obtained after the integrals over extra space coordinates. 
We, however, have shown by explicit calculation that $M_{H}^{2}$ is UV-finite, as is seen in (\ref{5.12}) or (\ref{6.8}). 
We thus conclude that in our model we do not suffer from divergent localized tadpole. 

We also point out that the ``global cancellation" of tadpoles, i.e. the cancellation among divergent localized tadpoles of 
physically distinct sectors of fixed points, does not happen either in our model. 
This is because the number of physically distinct sectors of fixed points is given by $[\frac{N}{2}]$ for $Z_{N}$ orbifold, 
 where $[\ldots]$ denotes the integer part \cite{SSSW}. 
Let us recall that the orbifold we are working on is $T_{2}/Z_{3}$ orbifold, which means that $[\frac{3}{2}] = 1$ and the global cancellation does not happen. 

Finally we mention the possible effect of brane localized gauge kinetic terms. 
If they ever exist they will cause, after the integral over the extra space coordinates, 
 the quantum corrections to the 4D kinetic terms of $h$ and $W^{\pm}$, 
 and the difference of the Wilson coefficients will affect $\Delta$ through wave-function renormalization. 
If the difference suffers from UV-divergence it may spoil the calculability of $\Delta$. 
(Concerning $\mu^{2}$, the effect of wave function renormalization affects the parameter $\mu^{2}$ only at the two-loop level and can be safely neglected in our analysis.) 
However, actually we do not suffer from the UV-divergence. 

The reasoning is similar to the one for the argument on the localized tadpole. 
We first point out that in the 6D bulk space-time, 
 the quantum corrections to the kinetic terms of the Higgs and $W^{\pm}$ boson are nothing but 
 those to the local operators $F_{\mu 5}F^{\mu 5}, \ F_{\mu 6}F^{\mu 6}$ and $F_{\mu \nu}F^{\mu \nu}$, 
 which are all included in a single operator $F_{MN}F^{MN}$. 
Thus the divergent parts of the quantum corrections to the kinetic terms of the Higgs and $W^{\pm}$ should be the same (see (\ref{7.17})). 
Hence, the only remaining possibility to suffer from the UV-divergence is due to the possible brane localized gauge kinetic terms, 
 which affects the kinetic terms in 4D effective theory after the integrals over extra space coordinates and therefore affects $\Delta$ through wave-function renormalization. 
We, however, have shown by explicit calculation that the predicted $\Delta$ including the contribution from the quantum corrections to the kinetic terms of $W^{\pm}_{\mu}$ 
 and $h_{0}$ is UV-finite, as is seen in (\ref{7.20}). Thus we conclude that the effect of the brane localized gauge kinetic terms, 
 even if they ever exist, should have been included in our analysis of $\Delta$ and do not spoil the calculability of the observable $\Delta$.

\subsection*{Acknowledgments}
This work was supported in part by the Grant-in-Aid for Scientific Research 
of the Ministry of Education, Science and Culture, Nos.~21244036,~23654090,~23104009 (C.S.L) and No.~24540283 (N.M.).


\end{document}